\documentclass[aps,prl,twocolumn,showpacs,preprintnumbers,superscriptaddress,dblfloatfix]{revtex4}
\usepackage{graphicx}   
\usepackage{dcolumn}   
\usepackage{bm}           
\usepackage{amssymb}  
\usepackage{setspace}
\usepackage{amsmath, amssymb, setspace}
\usepackage{array}
\usepackage{booktabs}
\usepackage{mathrsfs}
\usepackage{indentfirst}
\usepackage{slashed}
\usepackage{float}
\usepackage{lmodern}
\usepackage{hyperref}
\usepackage{xcolor}
\usepackage{multirow}
\usepackage{epstopdf}
\usepackage{ulem}
\usepackage{tabularx}
\usepackage[justification=raggedright]{caption}
\usepackage[flushleft]{threeparttable}

\begin{document}

\title{Search for Nucleon and Dinucleon Decays with an Invisible Particle and a Charged Lepton
in the Final State at the Super-Kamiokande Experiment}

\newcommand{\AFFicrr}{\affiliation{Kamioka Observatory, Institute for Cosmic Ray Research, University of Tokyo, Kamioka, Gifu 506-1205, Japan}}
\newcommand{\AFFkashiwa}{\affiliation{Research Center for Cosmic Neutrinos, Institute for Cosmic Ray Research, University of Tokyo, Kashiwa, Chiba 277-8582, Japan}}
\newcommand{\AFFipmu}{\affiliation{Kavli Institute for the Physics and
Mathematics of the Universe (WPI), Todai Institutes for Advanced Study,
University of Tokyo, Kashiwa, Chiba 277-8582, Japan }}
\newcommand{\AFFmad}{\affiliation{Department of Theoretical Physics, University Autonoma Madrid, 28049 Madrid, Spain}}
\newcommand{\AFFubc}{\affiliation{Department of Physics and Astronomy, University of British Columbia, Vancouver, BC, V6T1Z4, Canada}}
\newcommand{\AFFbu}{\affiliation{Department of Physics, Boston University, Boston, MA 02215, USA}}
\newcommand{\AFFbnl}{\affiliation{Physics Department, Brookhaven National Laboratory, Upton, NY 11973, USA}}
\newcommand{\AFFuci}{\affiliation{Department of Physics and Astronomy, University of California, Irvine, Irvine, CA 92697-4575, USA }}
\newcommand{\AFFcsu}{\affiliation{Department of Physics, California State University, Dominguez Hills, Carson, CA 90747, USA}}
\newcommand{\AFFcnm}{\affiliation{Department of Physics, Chonnam National University, Kwangju 500-757, Korea}}
\newcommand{\AFFduke}{\affiliation{Department of Physics, Duke University, Durham NC 27708, USA}}
\newcommand{\AFFfukuoka}{\affiliation{Junior College, Fukuoka Institute of Technology, Fukuoka, Fukuoka 811-0295, Japan}}
\newcommand{\AFFgifu}{\affiliation{Department of Physics, Gifu University, Gifu, Gifu 501-1193, Japan}}
\newcommand{\AFFgist}{\affiliation{GIST College, Gwangju Institute of Science and Technology, Gwangju 500-712, Korea}}
\newcommand{\AFFuh}{\affiliation{Department of Physics and Astronomy, University of Hawaii, Honolulu, HI 96822, USA}}
\newcommand{\AFFkek}{\affiliation{High Energy Accelerator Research Organization (KEK), Tsukuba, Ibaraki 305-0801, Japan }}
\newcommand{\AFFkobe}{\affiliation{Department of Physics, Kobe University, Kobe, Hyogo 657-8501, Japan}}
\newcommand{\AFFkyoto}{\affiliation{Department of Physics, Kyoto University, Kyoto, Kyoto 606-8502, Japan}}
\newcommand{\AFFmiyagi}{\affiliation{Department of Physics, Miyagi University of Education, Sendai, Miyagi 980-0845, Japan}}
\newcommand{\AFFnagoya}{\affiliation{Solar Terrestrial Environment Laboratory, Nagoya University, Nagoya, Aichi 464-8602, Japan}}
\newcommand{\AFFpol}{\affiliation{National Centre For Nuclear Research, 00-681 Warsaw, Poland}}
\newcommand{\AFFsuny}{\affiliation{Department of Physics and Astronomy, State University of New York at Stony Brook, NY 11794-3800, USA}}
\newcommand{\AFFokayama}{\affiliation{Department of Physics, Okayama University, Okayama, Okayama 700-8530, Japan }}
\newcommand{\AFFosaka}{\affiliation{Department of Physics, Osaka University, Toyonaka, Osaka 560-0043, Japan}}
\newcommand{\AFFregina}{\affiliation{Department of Physics, University of Regina, 3737 Wascana Parkway, Regina, SK, S4SOA2, Canada}}
\newcommand{\AFFseoul}{\affiliation{Department of Physics, Seoul National University, Seoul 151-742, Korea}}
\newcommand{\AFFshizuokasc}{\affiliation{Department of Informatics in
Social Welfare, Shizuoka University of Welfare, Yaizu, Shizuoka, 425-8611, Japan}}
\newcommand{\AFFskk}{\affiliation{Department of Physics, Sungkyunkwan University, Suwon 440-746, Korea}}
\newcommand{\AFFtokyo}{\affiliation{The University of Tokyo, Bunkyo, Tokyo 113-0033, Japan }}
\newcommand{\AFFtoronto}{\affiliation{Department of Physics, University of Toronto, 60 St., Toronto, Ontario, M5S1A7, Canada }}
\newcommand{\AFFtriumf}{\affiliation{TRIUMF, 4004 Wesbrook Mall, Vancouver, BC, V6T2A3, Canada }}
\newcommand{\AFFtokai}{\affiliation{Department of Physics, Tokai University, Hiratsuka, Kanagawa 259-1292, Japan}}
\newcommand{\AFFtsinghua}{\affiliation{Department of Engineering Physics, Tsinghua University, Beijing, 100084, China}}
\newcommand{\AFFuw}{\affiliation{Department of Physics, University of Washington, Seattle, WA 98195-1560, USA}}

\AFFicrr
\AFFkashiwa
\AFFmad
\AFFbu
\AFFubc
\AFFbnl
\AFFuci
\AFFcsu
\AFFcnm
\AFFduke
\AFFfukuoka
\AFFgifu
\AFFgist
\AFFuh
\AFFkek
\AFFkobe
\AFFkyoto
\AFFmiyagi
\AFFnagoya
\AFFpol
\AFFsuny
\AFFokayama
\AFFosaka
\AFFregina
\AFFseoul
\AFFshizuokasc
\AFFskk
\AFFtokai
\AFFtokyo
\AFFipmu
\AFFtoronto
\AFFtriumf
\AFFtsinghua
\AFFuw

\author{V.~Takhistov} 
\AFFuci

\author{K.~Abe}
\AFFicrr
\AFFipmu
\author{Y.~Haga}
\AFFicrr
\author{Y.~Hayato}
\AFFicrr
\AFFipmu
\author{M.~Ikeda}
\AFFicrr
\author{K.~Iyogi}
\AFFicrr 
\author{J.~Kameda}
\author{Y.~Kishimoto}
\author{M.~Miura} 
\author{S.~Moriyama} 
\author{M.~Nakahata}
\AFFicrr
\AFFipmu 
% ...

\collaboration{The Super-Kamiokande Collaboration}
\noaffiliation

\date{\today}

\begin{abstract}
Search results for nucleon decays $p \rightarrow e^+X$, $p \rightarrow \mu^+X$, 
$n \rightarrow \nu\gamma$
(where $X$ is an invisible, massless particle) as well as dinucleon decays 
$np \rightarrow e^+\nu$, $np \rightarrow \mu^+\nu$ and $np \rightarrow \tau^+\nu$ in the
Super-Kamiokande experiment are presented. 
Using single-ring data from an exposure of 
273.4 kton $\cdot$ years, a search
for these decays yields a result consistent with no signal. 
Accordingly, lower limits on the partial lifetimes of
$\tau_{p \rightarrow e^+X} > 7.9 \times 10^{32}$ years, 
$\tau_{p \rightarrow \mu^+X} > 4.1 \times 10^{32}$ years,
$\tau_{n \rightarrow \nu\gamma} > 5.5 \times 10^{32}$ years,
$\tau_{np \rightarrow e^+\nu} > 2.6 \times 10^{32}$ years,
$\tau_{np \rightarrow \mu^+\nu} > 2.2 \times 10^{32}$ years and
$\tau_{np \rightarrow \tau^+\nu} > 2.9 \times 10^{31}$ years
at a $90 \% $ confidence level are obtained.
Some of these searches are novel.
\end{abstract}



\maketitle

Many signs indicate that the Standard Model (SM) 
of particle physics is an incomplete description of nature.
Gauge coupling unification,
charge quantization and other features suggest a more unified account,
such as a Grand Unified Theory (GUT) 
\cite{Pati:1973uk,Pati:1973rp,Georgi:1974sy, Fritzsch:1974nn},
as an underlying fundamental theory. 
While the unification scale ($\sim 10^{15} - 10^{16}$ GeV)
is unreachable by accelerators, rare processes 
predicted by these theories, such as proton decay,
can be probed by large underground detectors. 
Being a signature prediction of GUTs, observation of an unstable
proton would constitute robust evidence for physics beyond the SM,
while non-observation will stringently constrain theoretical models.

The simplest unification scenarios based on minimal $\mathrm{SU}(5)$
and supersymmetric (SUSY) $\mathrm{SU}(5)$ have been decisively ruled out
by bounds on $p \rightarrow e^+\pi^0$ 
\cite{McGrew:1999nd, Hirata:1989kn, Shiozawa:1998si}
and $p \rightarrow \bar{\nu}K^+$ \cite{Kobayashi:2005pe}.
Many alternative scenarios as well as potential signatures are possible
(see \cite{Nath:2006ut} for review).

In this \textit{Letter}, we analyze a broad class of
nucleon and dinucleon decay channels with
a showering or non-showering single Cherenkov ring signature
 within the Super-Kamiokande (SK) experiment,
using the technique of spectral fit \cite{Fogli:2002pt, Takhistov:2014pfw}. 
First,  we have considered two general two-body decays
$p \rightarrow e^+X$ and $p \rightarrow \mu^+X$,
where $X$ is a single unknown invisible particle
which is assumed to be massless.
These searches are distinct from the
model-dependent inclusive analyses of \cite{Learned:1979gp, Cherry:1981uq}
listed in the PDG \cite{Agashe:2014kda}. Similarly,
we also consider $n \rightarrow \nu\gamma$.
Though this radiative process is suppressed,
it has a clean signature and has been considered
in the context of $\mathrm{SU}(5)$ \cite{Silverman:1980ha}, 
with some models \cite{Nath:2006ut}
predicting a lifetime of $10^{38 \pm 1}$ years.

While single nucleon $\Delta B = 1$
processes have been in general well studied, 
dinucleon $\Delta B = 2$ channels also
pose great interest. These higher-dimensional processes 
can become significant in models
which suppress proton decay and could be connected
to baryogenesis \cite{Arnold:2013cva}, accounting for the observed
baryon asymmetry of the universe \cite{Canetti:2012zc}.
Such a connection may already be hinted at from the 
requirement of baryon number violation as a necessary condition
for explaining the asymmetry \cite{Sakharov:1967dj}.
The disappearance $\Delta B = 2$ reactions,
with invisible final state particles,
 have been studied and no signal excess was 
observed \cite{Berger:1991fa,Araki:2005jt, Litos:2014fxa}.
The channels $np \rightarrow e^+\nu$, $np \rightarrow \mu^+\nu$
 and  $np \rightarrow \tau^+\nu$
violate baryon number by two units and violate lepton number
by either two or zero units. They can become significant in 
models with an extended Higgs sector \cite{Mohapatra:1982aj, Arnold:2013cva},
which could be considered in the context of GUTs \cite{Arnellos:1982nt}.
While the $\tau$ cannot occur in single nucleon decay,
in dinucleon decay the $\tau$ channel is allowed \cite{Bryman:2014tta}.
The process $np \rightarrow \tau^+\nu$ has not been experimentally studied before
and in addition to the electron and muon channel searches
we present the first search in the $\tau$ channel.
 
In this work, SK data is analyzed from an exposure of the 22.5 kton fiducial mass for
273.4 kton$\cdot$years, covering four running periods (SK-I through SK-IV).
Details of the detector design and performance in each SK period,
can be found in~\cite{Fukuda:2002uc,Abe:2013gga}.~This
 analysis considers only events in which all 
observed Cherenkov light was fully contained (FC) within the inner detector.

Since final-state neutrinos or $X$ (by definition) 
are not observed, the only signature of $p \rightarrow e^+X$, $p \rightarrow \mu^+X$, 
$n \rightarrow \nu\gamma$, $np \rightarrow e^+\nu$ and $np \rightarrow \mu^+\nu$ 
is a single charged $e^+$ or $\mu^+$ lepton, or single $\gamma$. 
Thus, the invariant mass of the initial state cannot be reconstructed
and the signal will be superimposed on a substantial
atmospheric neutrino background in the $e$-like and
$\mu$-like momentum spectra. For the $np \rightarrow \tau^+\nu$ decay, only the
$\tau \rightarrow e^+\nu\nu$ and $\tau \rightarrow \mu^+\nu\nu$ channels
are considered, with the respective branching ratios of 17.8\% and 17.4\%. 
This allows us to perform all the analyses within a unified framework.
The previous searches for $n \rightarrow \nu\gamma$, 
$np \rightarrow e^+\nu$ and $np \rightarrow \mu^+\nu$,
which were performed with a smaller detector using a counting method,
resulted in the lifetime limits of $2.8 \times 10^{31}$ years \cite{McGrew:1999nd},
$2.8 \times 10^{30}$ years \cite{Berger:1991fa} and
$1.6 \times 10^{30}$ years \cite{Berger:1991fa}, respectively.
In contrast, the spectral fit employed within this work, allows
utilization of the extra information from the 
energy dependence of signal, background and the systematic errors.
 This methodology has been recently employed in 
the SK nucleon decay \cite{Abe:2013lua,Takhistov:2014pfw}
and dark matter analyses \cite{Choi:2015ara}.

The nucleon decay signal events are obtained
from Monte Carlo (MC) simulations, in which
all the nucleons of the water H$_2$O molecule are assumed
to decay with equal probability. The final state particles are generated 
with energy and momentum uniformly distributed in phase space.
The effects of Fermi motion, nuclear binding energy as well
as nucleon - nucleon correlated decays \cite{Yamazaki:1999gz} 
are taken into account for both nucleon \cite{Nishino:2012ipa, Regis:2012sn,Abe:2014mwa}
and dinucleon searches \cite{Gustafson:2015qyo}. 
The signal Fermi momentum distributions are simulated using a
spectral function fit to electron-$^{12}$C scattering data \cite{Nakamura:1976mb}.
The SK detector simulation~\cite{Abe:2013gga} is based on the GEANT-3 \cite{Brun:1994aa} package,
with the TAUOLA \cite{Jadach:1993hs} package employed for decaying the $\tau$ leptons.
For the $np \rightarrow \tau^+\nu$ mode we generated three MC samples, 
with the $\tau$ decaying to $e^+\nu\nu$, 
$\mu^+\nu\nu$ and all decay channels.
The latter allows us to study 
sample contamination in the two selected
leptonic $\tau$ channels from the hadronic $\tau$
channels and thus identify sample purity after the event selection.
We have confirmed that the resulting MC charged
lepton spectra from $\tau \rightarrow e^+\nu\nu$ and $\tau \rightarrow \mu^+\nu\nu$ decays
agree with the theoretical formula \cite{Chen:2014ifa}. 
For $p \rightarrow e^+X$ and $p \rightarrow \mu^+X$ modes,
the invisible $X$ particle cannot be a fermion by spin conservation, 
but in our spin-insensitive MC it was simulated as a neutrino.
In total, around 4,200 signal events were generated within the fiducial volume (FV) 
for each SK period for single nucleon decays and around 8,400 for dinucleon decays.

Atmospheric neutrino background interactions were generated 
using the flux of Honda \textit{et.~al.}~\cite{Honda:2006qj}
and the NEUT simulation package~\cite{Hayato:2002sd},
which uses a relativistic Fermi gas model.
Background MC corresponding to a 500-year
exposure of the detector was simulated for each detector phase.
We used the same atmospheric neutrino MC
 as the standard SK oscillation analysis \cite{Abe:2014gda}.

The event selection applied to the fully-contained data is the following:
(A) a single Cherenkov ring is present,
(B) the ring is showering (electron-like) for $p \rightarrow  e^+X$,
$n \rightarrow \nu\gamma$, $np \rightarrow e^+\nu$ and 
$np \rightarrow \tau^+\nu~(\tau \rightarrow e^+\nu\nu)$
 and non-showering (muon-like) for $p \rightarrow  \mu^+X$,
$np \rightarrow \mu^+\nu$ and $np \rightarrow \tau^+\nu~(\tau \rightarrow \mu^+\nu\nu)$,
(C) there are zero decay electrons for modes with an $e$-like ring
 and one decay electron for those with a $\mu$-like ring,
(D) the reconstructed momentum lies in the range 100 MeV/$c$ $ \le p_e \le $ 1000 MeV/$c$
 for $p \rightarrow  e^+X$, $n \rightarrow \nu\gamma$ 
and in the range 200 MeV/$c$ $ \le p_\mu \le $ 1000 MeV/$c$
 for $p \rightarrow \mu^+X$, with the range extended 
to 100-1500 MeV/$c$ for dinucleon decays with an $e$-like ring
and 200-1500 MeV/$c$ for those with a $\mu$-like ring.
In total, approximately 37,000 FC events were obtained in the
SK-I to SK-IV data-taking periods.
After the criteria (A)-(D) have been applied,
the final data samples for single nucleon decay searches with an $e$-like ring
contain 8,500 events and 6,000 events for the case of $\mu$-like ring,
with momenta up to one GeV/c. To search for dinucleon 
decays we consider lepton momenta up to 1500 MeV/c.
The final samples for the dinucleon modes contain
9,500 events for the $e$-like channels and 6,500 events for the $\mu$-like ones.
See Ref. \cite{Shiozawa:1999sd} for details regarding reconstruction.

\begin{table*}[t]
\caption{Systematic errors of spectrum fits, with $1\sigma$
uncertainties and resulting fit pull terms. 
Errors specific to signal and background are denoted by S and B, 
while those that are common to both by SB.}  
\begin{threeparttable}[htb]
\begin{tabular}{lccccc} 
\hline 
\hline
   		Decay mode			 & ~ & $p \rightarrow e^+X$ &  $p \rightarrow \mu^+X$ &\\
\hline
    Systematic error  & 	1-$\sigma$ uncertainty (\%) & Fit pull ($\sigma$)	 & Fit pull ($\sigma$)  & \\
    Final state interactions (FSI)								& 10  				    &  ~0.10		 				&  -0.60		     				& ~B \\
    Flux normalization ($E_\nu < 1$ GeV)  					& ~25 \tnote{a}        &  -0.23		 					&  -0.08 		     				& ~B \\
    Flux normalization ($E_\nu > 1$ GeV)  					& ~15 \tnote{b}        &  -1.44	  						&  -0.50						& ~B \\
    $M_A$ in $\nu$ interactions							     &	10  		  	    &  ~0.69		 				&  ~0.23 		               	     & ~B \\
    Single meson cross-section in $\nu$ interactions			&	10  			    &  -0.55  		   				&  -0.14  					     & ~B \\
    Energy calibration of SK-I, -II, -III, -IV          				& 1.1, 1.7, 2.7, 2.3      & 0.58, -0.91, 0.48, 0.38~~	& -0.54, 0.07, -0.14, 0.26		& ~SB \\
    Fermi model comparison         								& ~10 \tnote{c}	     &  -0.08  	  		 			& ~0.70						& ~S \\		
    Nucleon-nucleon correlated decay         					&  100  				& ~0.00		   				& ~0.06				           & ~S \\
\hline 
\hline
\end{tabular} 
\begin{tablenotes}[flushleft]\footnotesize
 \item [a] Uncertainty linearly decreases with $\log{E_\nu}$ from 25\% (0.1 GeV) to 7\% (1 GeV). \\
 \item [b] Uncertainty is 7\% up to 10 GeV, linearly increases with $\log E_{\nu}$ from 7\% (10 GeV)
               to 12\% (100 GeV) and then 20\% (1 TeV). \\
 \item [c] Estimated from comparison of spectral function and Fermi gas model.
\end{tablenotes}
\end{threeparttable}
\label{tab:syserr}
\end{table*}

The signal detection efficiency is defined as the fraction of events passing selection criteria
compared to the total number of events 
generated within the true fiducial volume.
The average detection efficiency for $e$-like channels is
$94.0 \pm 0.4 \%$ for all SK data-taking periods.
For the $\mu$-like channels, the average 
detection efficiency is $76.4 \pm 0.6 \%$ for SK-I to SK-III
and $91.7 \pm 0.4 \%$ for SK-IV.
The increase in efficiency observed in SK-IV for 
channels with a $\mu$-like ring, comes from a 20\% improvement in
the detection of muon decay electrons after
 an upgrade of the detector electronics \cite{Abe:2013gga}.

For the $e$-like momentum spectrum up to 1500 MeV/c,
the dominant background contribution, composing $75.8\%$ of the events, 
comes from the $\nu_e$ charged-current (CC) quasi-elastic (QE) neutrino channel.
The $\nu_e$ CC single-pion production constitutes $13.0\%$ of the background,
while the $\nu_e$ CC coherent-pion, 
CC multi-pion and neutral-current (NC) single-pion
productions contribute around $1.1\%$, $1.1\%$ and $1.6\%$, respectively.
About $3.5\%$ and $1.1\%$ of events come from $\nu_{\mu}$
NC single-pion and coherent-pion production. For the $\mu$-like
momentum spectrum up to 1500 MeV/c, the dominant contribution
of around $78.6\%$ comes from $\nu_{\mu}$ CCQE.
Similarly, $\nu_{\mu}$ CC single-pion, CC coherent-pion 
and CC multi-pion as well as NC single-pion production contribute
around $16.2\%$, $1.4\%$, $1.6\%$ and $0.8\%$, respectively.

After event selection, a spectral
fit is performed on the reconstructed charged lepton
momentum distribution of the events.
The $\chi^2$ minimization fit is based on the Poisson distribution, 
with the systematic uncertainties accounted 
for by quadratic penalties (``pull terms") \cite{Fogli:2002pt}.
The $\chi^2$ function used in the analysis is
\begin{equation}
\begin{split}
 \chi^{2} = ~& 2 \sum^{{\textrm{nbins}}}_{i=1}  \Big( N^{\textrm{exp}}_i + N^{{\textrm{obs}}}_{i} 
		\Big[ \ln \frac{N^{{\textrm{obs}}}_i}{ N^{\textrm{exp}}_i} -1 \Big] \Big)
             + \sum^{N_{{\textrm{syserr}}}}_{j=1} ( \frac{ \epsilon_{j} }{ \sigma_{j} } )^{2}  \\
& N^{\textrm{exp}}_i = \Big[ \alpha \cdot N^{{\textrm{back}}}_{i} + \beta \cdot N^{{\textrm{sig}}}_{i} \Big]
 ( 1 + \sum^{N_{{\textrm{syserr}}}}_{j=1} f^{j}_{i} \frac{ \epsilon_{j} }{ \sigma_{j}}  ) , 
\end{split}
\label{eq:chi}
\end{equation}
where $i$ labels the analysis bin. 
The terms $N^{{\textrm{obs}}}_{i}$, $N^{{\textrm{sig}}}_{i}$, $N^{{\textrm{back}}}_{i}$,
$N^{{\textrm{exp}}}_{i}$ are the numbers of observed data, signal MC, background MC

\begin{table*}[htb]
\caption{Best fit $(\alpha, \beta)$ parameter values, best fit $\chi^2 /$ d.o.f.,
no signal $\Delta \chi^2$, 90\%~C.L. value of $\beta$ parameter,
allowed number of nucleon decay events in the full 273.4~kton $\cdot$ years
exposure and a partial lifetime limit for each decay mode at 90\% C.L. 
The sensitivity and lifetime limit for dinucleon decay modes are per $^{16}$O nucleus.}  
\begin{tabularx}{\textwidth}{X *{7}{>{\centering\arraybackslash}X}}
\hline 
\hline
   Decay mode   &  Best fit  & Best fit & No signal &  Data &  Data &  Sensitivity & $\tau/{\mathcal{B}}$ \\ 
\hline
               &  $(\alpha,\beta)$  & $\chi^2 / \text{d.o.f.}$ & $\Delta \chi^2$   & $\beta_{{\mathrm{90CL}}}$ & $N_{{\mathrm{90CL}}}$ & ($\times10^{31}$~yr.) & ($\times10^{31}$~yr.) \\
   $p \rightarrow  e^+X$					& (1.050, 0.002)		& 70.9/70		& 0.19			& 0.013 	& 108 	& 79 			& 79 \\	
   $n \rightarrow \nu\gamma$	 			& (1.045, 0.004) 		& 70.5/70		& 0.43			& 0.015 	& 125 	& 58			& 55 \\			
   $p \rightarrow \mu^+ X$		 		& (0.960, 0.016)  		& 63.2/62 		& 3.43			& 0.032 	& 187 	& 77			& 41 \\
   $np \rightarrow e^+\nu$	 			& (0.955, 0.000)  		& 122.5/110	& 0.00		 	& 0.004 	& 33 	& 10			& 26 \\	
   $np \rightarrow \mu^+\nu$			& (0.910, 0.000) 		& 97.0/102		& 0.00		 	& 0.005 	& 36 	& 11			& 20 \\
   $np \rightarrow \tau^+\nu$	 		& (0.910, 0.000)    	& 224.6/214  	& 0.00		 	& 0.006 	& 96	 & 1			& 3 \\
\hline 
\hline 
\end{tabularx} 
\label{tab:results}
\vspace{-1em}
\end{table*}

and the total (signal and background) MC events in each bin $i$. 
The index $j$ labels the systematic errors, while $\epsilon_{j}$ and $f_{i}^{j}$ 
correspond to the fit error parameter
 and the fractional change in the $N^{{\textrm{exp}}}_{i}$
bin due to 1-sigma error uncertainty $\sigma_{j}$, respectively.
The fit is performed for two parameters $\alpha$ and $\beta$,
which denote the background and signal normalizations, respectively.
After the event selection, the signal MC 
distribution is normalized to the background by the integral,
which in turn is normalized to the SK livetime.
This allows us to identify the fit point $(\alpha, \beta) = (1, 0)$
with the no-signal hypothesis.
Similarly, $(\alpha, \beta) = (0, 1)$ signifies that
the data is described by signal only, with the signal
amount equal to background MC normalized (pre-fit) to livetime.
The $\chi^2$ minimization is carried out 
over each $\alpha$ and $\beta$ in the grid
according to $ \partial \chi^2/  \partial \epsilon_{j} = 0$.
The resulting global minimum is defined as the best fit.
Further details on the fit and specifics of systematic error treatment can
be found in \cite{Wendell:2010md, Takhistov:2014pfw, Choi:2015ara}.
For the $np \rightarrow \tau^+\nu\nu$ mode, 
after the appropriate event selection is applied to both
MC samples of $\tau \rightarrow e^+\nu\nu$ and $\tau \rightarrow \mu^+\nu\nu$,
the samples are combined for the fit, allowing us to obtain a single value for
the permitted number of nucleon decays at 90\% CL.

The systematic errors can be divided into signal-specific (S), 
background-specific (B) as well as detector and reconstruction errors, 
which are common to both signal and background (SB). 
The two signal specific systematics are from Fermi motion 
and nucleon - nucleon correlated decay. For background, 
in order to methodically select the dominant systematics, 
we started from more than 150 errors employed in 
the SK oscillation analysis \cite{Wendell:2010md} 
and chose those which affect the analyses bins by more than 5\% 
($|f_i^j| \ge 0.05$). Relaxing this criteria to 1\% does not significantly
alter the results, but complicates the analysis \cite{Takhistov:2014pfw}.~As
 in \cite{Takhistov:2014pfw}, we have found that the dominant 
contributions originate from uncertainties related to neutrino flux and
energy calibration (common to both signal and background). 
Including the signal systematics, the total number of considered 
errors is 11 and they are the same for all modes. 
In Table~\ref{tab:syserr} we display the complete list of systematics, 
their uncertainties and fitted pull terms for two representative 
examples $p \rightarrow e^+X$ and $p \rightarrow \mu^+X$.

The spectral fit determines the overall background 
and signal normalizations $\alpha$ and $\beta$,
with the fit results displayed in Table~\ref{tab:results}.
The outcome shows that no significant signal excess has been observed,
with the data $\Delta \chi^2 = \chi^2 - \chi_{min}^2$ being within
$1 \sigma$ of the background only hypothesis for all search modes except 
for $p \rightarrow \mu^+ X$, which is within $2 \sigma$.

\begin{figure*}[t]
\begin{minipage}[b]{0.45\textwidth}
\centering
\hspace{-3em}
\includegraphics[width=\textwidth]{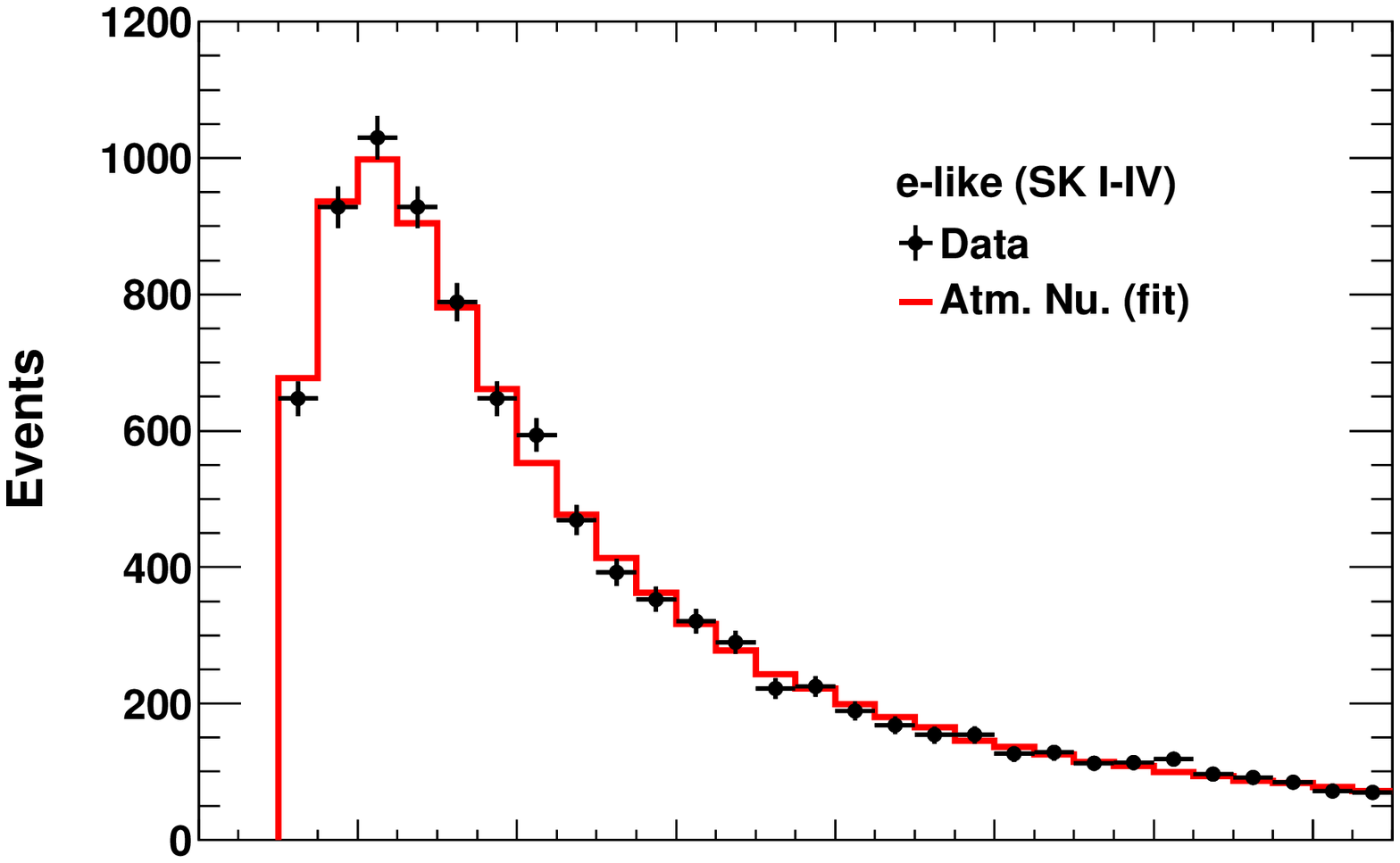}
\end{minipage}
\begin{minipage}[b]{0.45\textwidth}
\centering
\hspace{-3em}
\includegraphics[width=\textwidth]{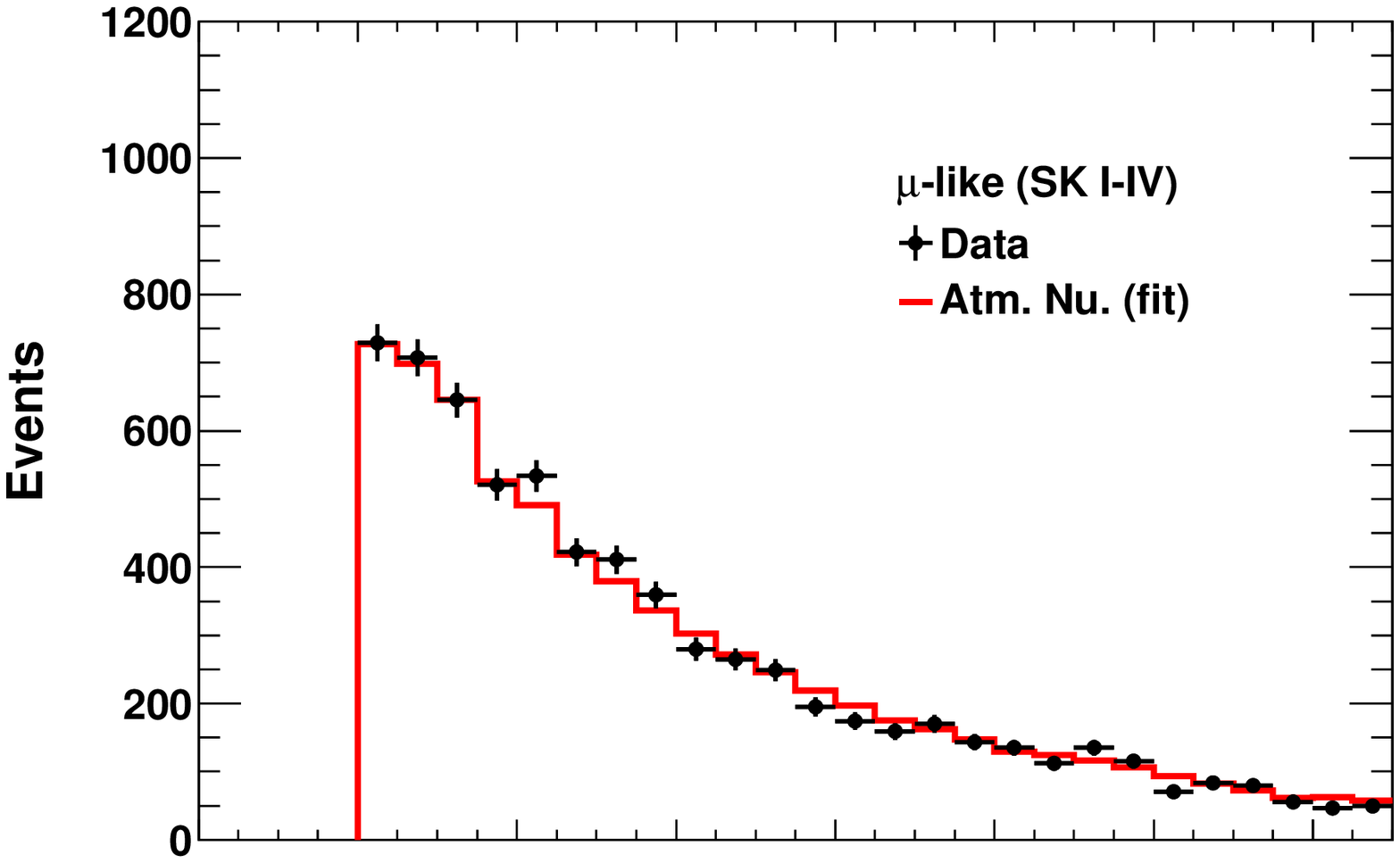}
\end{minipage}
\begin{minipage}[b]{0.45\textwidth}
\centering
\hspace{-3em}
\includegraphics[width=\textwidth]{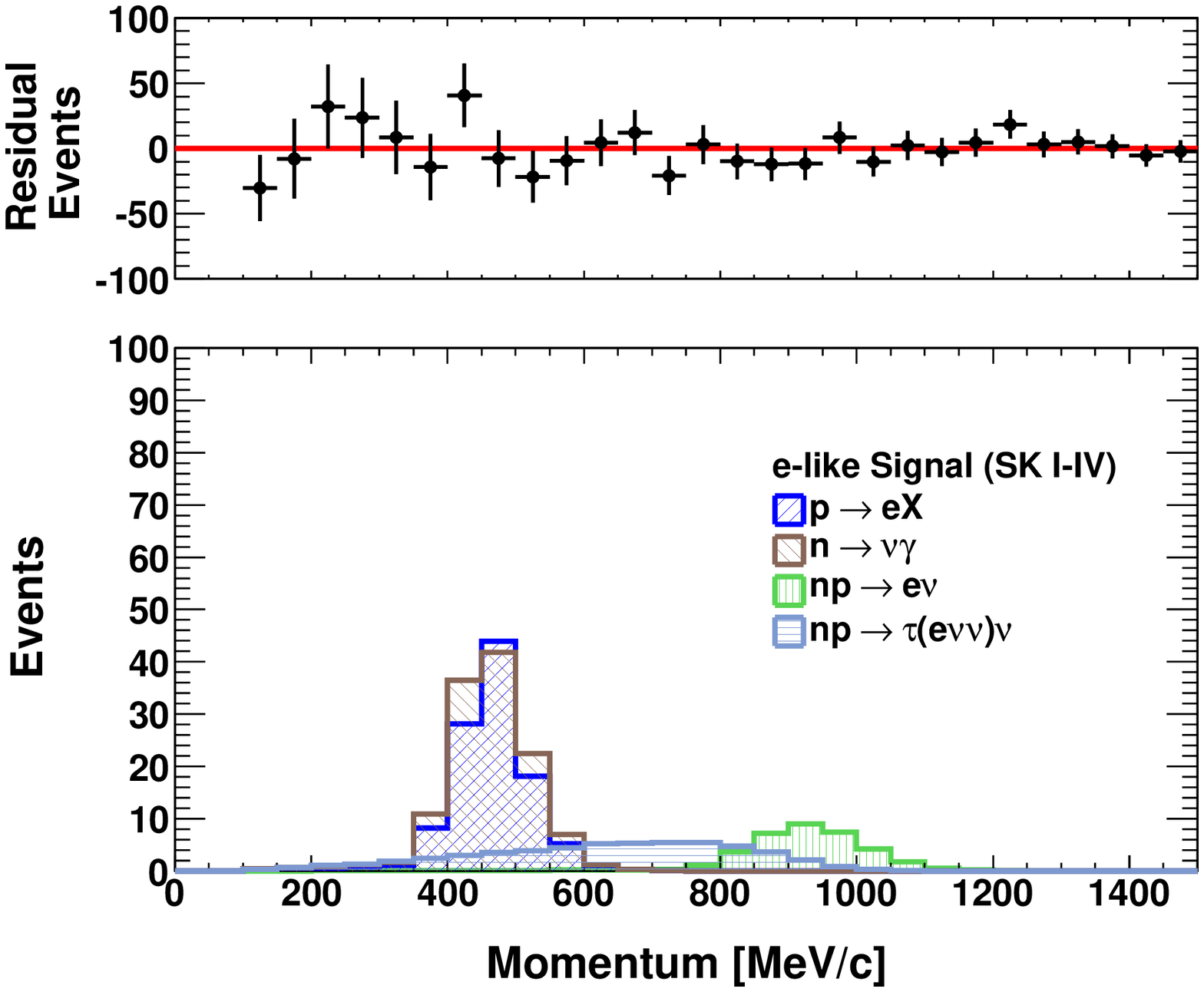}
\end{minipage}
\begin{minipage}[b]{0.45\textwidth}
\centering
\hspace{-3em}
\includegraphics[width=\textwidth]{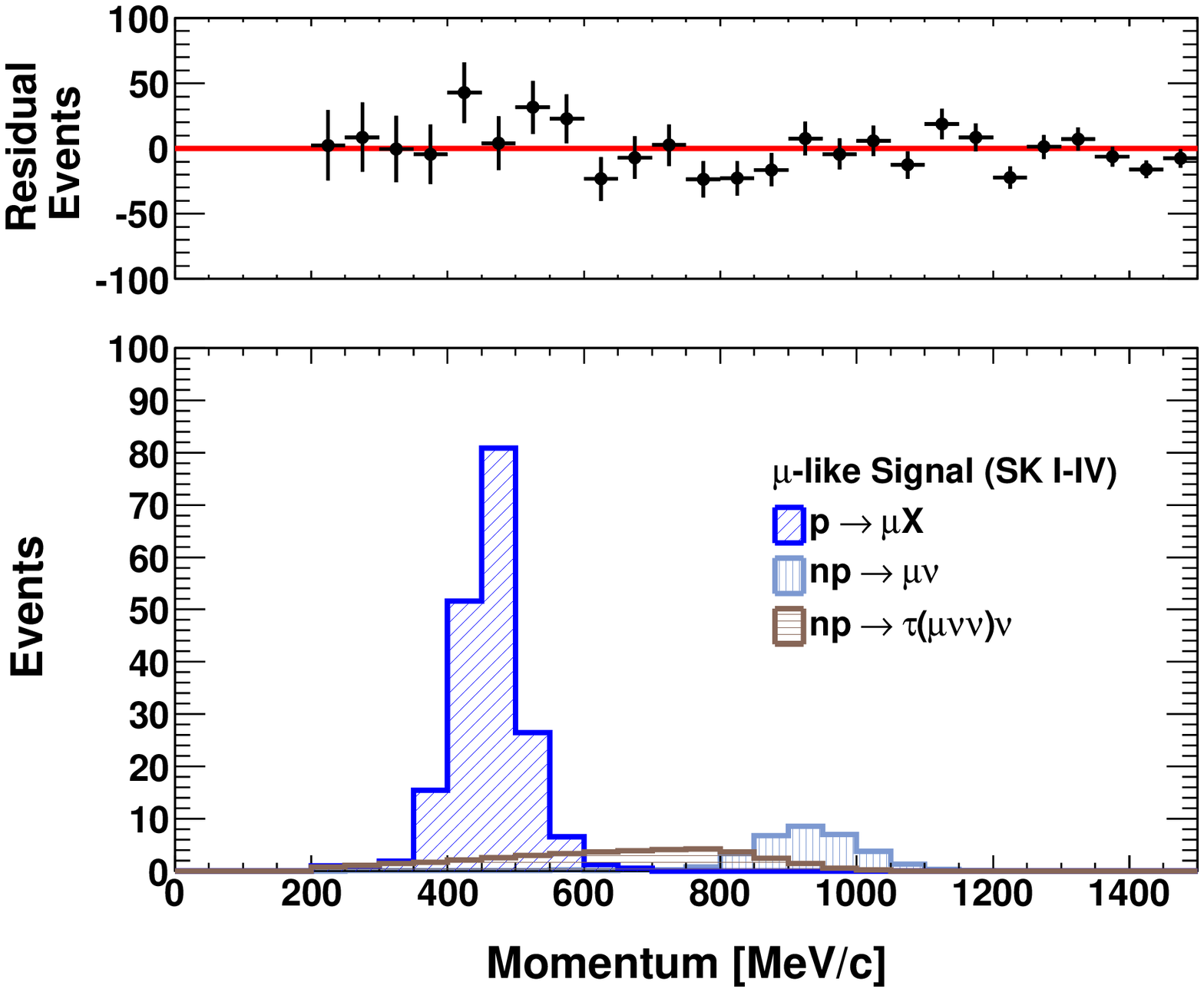}
\end{minipage}

\caption{[Top] Reconstructed momentum distribution for 273.4 kton $\cdot$ years 
of combined SK data (black dots) and the best-fit
result for the atmospheric neutrino background Monte Carlo (solid line).
The corresponding residuals are shown below, after fitted background subtraction from data.
[Bottom] The 90\% confidence level allowed nucleon decay signal (hatched histograms),
from the signal and background MC fit to data. All modes are shown (overlaid),
with $e$-like channels on the left and $\mu$-like channels on the right.}
\label{fig:results_full}
\end{figure*}
The lower lifetime limit on the processes can then be computed from the
90\% confidence level value of $\beta$ ($\beta_{90 \text{CL}}$),
which translates into the allowed amount of signal at 90\% confidence level
according to $N_{90 \text{CL}} = \beta_{90 \text{CL}} \cdot N^{\text{signal}}$,
where $N^{\text{signal}}$ is the total number of signal events.
The partial lifetime limit is then calculated from
\begin{equation}
	\tau_{90 \text{CL}} /\mathcal{B} ~=~\frac{\sum^{{\textrm{SK4}}}_{\text{sk} 
=\textrm{SK1}}\lambda_{\text{sk}} \cdot \epsilon_{\text{sk}} \cdot  N^{\text{nucleons}}}{N_{90\text{CL}}},
\end{equation}
\noindent where $\mathcal{B}$ is the branching ratio of a process,
$\epsilon_{\text{sk}}$ and $\lambda_{\text{sk}}$ are the signal efficiency
and the exposure in kton $\cdot$ years for each SK phase,
$N_{90\text{CL}}$ is the amount of signal allowed at the 90\%
confidence level and $N^{\text{nucleons}}$ is the number of nucleons per kiloton of water,
corresponding to $3.3 \times 10^{32}$, $2.7 \times 10^{32}$
and $3.3 \times 10^{31}$ for proton, neutron and dinucleon decay
searches, respectively.

The resulting fitted spectra for the 273.4 kton $\cdot$ years 
of combined SK data can be found in Figure \ref{fig:results_full}.
The upper figures display best-fit result for 
atmospheric neutrino background (solid line) without signal
fitted to data (black dots) and the
corresponding residuals after the fitted MC is subtracted from data.
It is seen that the background MC describes the data well.
The bottom figures display the 90\% C.L. allowed signal (hatched histogram),
 obtained from the fit of background with signal to data,
with all the $e$-like and $\mu$-like spectra overlaid with all the modes.
The $N_{90CL}$ as well as resulting sensitivities 
and calculated lifetime limits for the decays
are shown in Table~\ref{tab:results}. The sensitivities
were obtained assuming that data are described by background.
For the $np \rightarrow \tau\nu$ mode we have combined 
the $\tau$ channels $e^+\nu\nu$ and $\mu^+\nu\nu$,
weighted by their respective branching ratios. 
This limit is then multiplied by 1.15 to account for roughly $85 \%$
sample purity of the tau channels.
We set the lower limits on the partial lifetimes of the decay modes
at the $90 \%$ C.L., with the results shown in~Table~\ref{tab:results}.

In conclusion, the single Cherenkov ring momentum spectra in Super-Kamiokande
are well described by atmospheric neutrinos, including the effect of neutrino oscillation
 and systematic uncertainties, up to 1500 MeV/$c$. 
We find no evidence for any contribution from the
six different nucleon and dinucleon decay modes 
that would produce a showering or non-showering Cherenkov ring.
The results of this analysis provide a stringent test of new physics.
The obtained limits represent more than an order of
magnitude improvement over the previous analyses of
$n \rightarrow \nu\gamma$ \cite{McGrew:1999nd} and two orders of magnitude
for $np \rightarrow e^+\nu$ and $np \rightarrow \mu^+\nu$ \cite{Berger:1991fa}.
The searches for $p \rightarrow e^+X$, 
$p \rightarrow \mu^+X$ (where $X$ is an invisible,
massless particle) and $np \rightarrow \tau^+\nu$ are novel. 
The dinucleon decay limits restrict $\Delta B = 2$ processes with $L$ violated by
either zero or two units.

\indent 
We gratefully acknowledge the cooperation of the
Kamioka Mining and Smelting Company.~The Super-Kamiokande
experiment was built and has been operated
with funding from the Japanese Ministry of Education,
Culture, Sports, Science and Technology, the U.S. Department
of Energy, and the U.S. National Science Foundation.
%

\bibliography{nucleonbib}

\begin{thebibliography}{42}
\expandafter\ifx\csname natexlab\endcsname\relax\def\natexlab#1{#1}\fi
\expandafter\ifx\csname bibnamefont\endcsname\relax
  \def\bibnamefont#1{#1}\fi
\expandafter\ifx\csname bibfnamefont\endcsname\relax
  \def\bibfnamefont#1{#1}\fi
\expandafter\ifx\csname citenamefont\endcsname\relax
  \def\citenamefont#1{#1}\fi
\expandafter\ifx\csname url\endcsname\relax
  \def\url#1{\texttt{#1}}\fi
\expandafter\ifx\csname urlprefix\endcsname\relax\def\urlprefix{URL }\fi
\providecommand{\bibinfo}[2]{#2}
\providecommand{\eprint}[2][]{\url{#2}}

\bibitem[{\citenamefont{Pati and Salam}(1973{\natexlab{a}})}]{Pati:1973uk}
\bibinfo{author}{\bibfnamefont{J.~C.} \bibnamefont{Pati}} \bibnamefont{and}
  \bibinfo{author}{\bibfnamefont{A.}~\bibnamefont{Salam}},
  \bibinfo{journal}{Phys. Rev.} \textbf{\bibinfo{volume}{D8}},
  \bibinfo{pages}{1240} (\bibinfo{year}{1973}{\natexlab{a}}).

\bibitem[{\citenamefont{Pati and Salam}(1973{\natexlab{b}})}]{Pati:1973rp}
\bibinfo{author}{\bibfnamefont{J.~C.} \bibnamefont{Pati}} \bibnamefont{and}
  \bibinfo{author}{\bibfnamefont{A.}~\bibnamefont{Salam}},
  \bibinfo{journal}{Phys. Rev. Lett.} \textbf{\bibinfo{volume}{31}},
  \bibinfo{pages}{661} (\bibinfo{year}{1973}{\natexlab{b}}).

\bibitem[{\citenamefont{Georgi and Glashow}(1974)}]{Georgi:1974sy}
\bibinfo{author}{\bibfnamefont{H.}~\bibnamefont{Georgi}} \bibnamefont{and}
  \bibinfo{author}{\bibfnamefont{S.}~\bibnamefont{Glashow}},
  \bibinfo{journal}{Phys.Rev.Lett.} \textbf{\bibinfo{volume}{32}},
  \bibinfo{pages}{438} (\bibinfo{year}{1974}).

\bibitem[{\citenamefont{Fritzsch and Minkowski}(1975)}]{Fritzsch:1974nn}
\bibinfo{author}{\bibfnamefont{H.}~\bibnamefont{Fritzsch}} \bibnamefont{and}
  \bibinfo{author}{\bibfnamefont{P.}~\bibnamefont{Minkowski}},
  \bibinfo{journal}{Annals Phys.} \textbf{\bibinfo{volume}{93}},
  \bibinfo{pages}{193} (\bibinfo{year}{1975}).

\bibitem[{\citenamefont{McGrew et~al.}(1999)\citenamefont{McGrew,
  Becker-Szendy, Bratton, Breault, Cady et~al.}}]{McGrew:1999nd}
\bibinfo{author}{\bibfnamefont{C.}~\bibnamefont{McGrew}},
  \bibinfo{author}{\bibfnamefont{R.}~\bibnamefont{Becker-Szendy}},
  \bibinfo{author}{\bibfnamefont{C.}~\bibnamefont{Bratton}},
  \bibinfo{author}{\bibfnamefont{J.}~\bibnamefont{Breault}},
  \bibinfo{author}{\bibfnamefont{D.}~\bibnamefont{Cady}}, \bibnamefont{et~al.}
  (\bibinfo{collaboration}{IMB-3 Collaboration}), \bibinfo{journal}{Phys.Rev.}
  \textbf{\bibinfo{volume}{D59}}, \bibinfo{pages}{052004}
  (\bibinfo{year}{1999}).

\bibitem[{\citenamefont{Hirata et~al.}(1989)}]{Hirata:1989kn}
\bibinfo{author}{\bibfnamefont{K.}~\bibnamefont{Hirata}} \bibnamefont{et~al.}
  (\bibinfo{collaboration}{KAMIOKANDE-II Collaboration}),
  \bibinfo{journal}{Phys.Lett.} \textbf{\bibinfo{volume}{B220}},
  \bibinfo{pages}{308} (\bibinfo{year}{1989}).

\bibitem[{\citenamefont{Shiozawa et~al.}(1998)}]{Shiozawa:1998si}
\bibinfo{author}{\bibfnamefont{M.}~\bibnamefont{Shiozawa}} \bibnamefont{et~al.}
  (\bibinfo{collaboration}{Super-Kamiokande Collaboration}),
  \bibinfo{journal}{Phys.Rev.Lett.} \textbf{\bibinfo{volume}{81}},
  \bibinfo{pages}{3319} (\bibinfo{year}{1998}), \eprint{hep-ex/9806014}.

\bibitem[{\citenamefont{Kobayashi et~al.}(2005)}]{Kobayashi:2005pe}
\bibinfo{author}{\bibfnamefont{K.}~\bibnamefont{Kobayashi}}
  \bibnamefont{et~al.} (\bibinfo{collaboration}{Super-Kamiokande
  Collaboration}), \bibinfo{journal}{Phys.Rev.} \textbf{\bibinfo{volume}{D72}},
  \bibinfo{pages}{052007} (\bibinfo{year}{2005}), \eprint{hep-ex/0502026}.

\bibitem[{\citenamefont{Nath and Fileviez~Perez}(2007)}]{Nath:2006ut}
\bibinfo{author}{\bibfnamefont{P.}~\bibnamefont{Nath}} \bibnamefont{and}
  \bibinfo{author}{\bibfnamefont{P.}~\bibnamefont{Fileviez~Perez}},
  \bibinfo{journal}{Phys.Rept.} \textbf{\bibinfo{volume}{441}},
  \bibinfo{pages}{191} (\bibinfo{year}{2007}), \eprint{hep-ph/0601023}.

\bibitem[{\citenamefont{Fogli et~al.}(2002)\citenamefont{Fogli, Lisi, Marrone,
  Montanino, and Palazzo}}]{Fogli:2002pt}
\bibinfo{author}{\bibfnamefont{G.}~\bibnamefont{Fogli}},
  \bibinfo{author}{\bibfnamefont{E.}~\bibnamefont{Lisi}},
  \bibinfo{author}{\bibfnamefont{A.}~\bibnamefont{Marrone}},
  \bibinfo{author}{\bibfnamefont{D.}~\bibnamefont{Montanino}},
  \bibnamefont{and} \bibinfo{author}{\bibfnamefont{A.}~\bibnamefont{Palazzo}},
  \bibinfo{journal}{Phys.Rev.} \textbf{\bibinfo{volume}{D66}},
  \bibinfo{pages}{053010} (\bibinfo{year}{2002}), \eprint{hep-ph/0206162}.

\bibitem[{\citenamefont{Takhistov et~al.}(2014)}]{Takhistov:2014pfw}
\bibinfo{author}{\bibfnamefont{V.}~\bibnamefont{Takhistov}}
  \bibnamefont{et~al.} (\bibinfo{collaboration}{Super-Kamiokande
  Collaboration}), \bibinfo{journal}{Phys.Rev.Lett.}
  \textbf{\bibinfo{volume}{113}}, \bibinfo{pages}{101801}
  (\bibinfo{year}{2014}), \eprint{1409.1947}.

\bibitem[{\citenamefont{Learned et~al.}(1979)\citenamefont{Learned, Reines, and
  Soni}}]{Learned:1979gp}
\bibinfo{author}{\bibfnamefont{J.}~\bibnamefont{Learned}},
  \bibinfo{author}{\bibfnamefont{F.}~\bibnamefont{Reines}}, \bibnamefont{and}
  \bibinfo{author}{\bibfnamefont{A.}~\bibnamefont{Soni}},
  \bibinfo{journal}{Phys.Rev.Lett.} \textbf{\bibinfo{volume}{43}},
  \bibinfo{pages}{907} (\bibinfo{year}{1979}).

\bibitem[{\citenamefont{Cherry et~al.}(1981)\citenamefont{Cherry, Deakyne,
  Lande, Lee, Steinberg et~al.}}]{Cherry:1981uq}
\bibinfo{author}{\bibfnamefont{M.}~\bibnamefont{Cherry}},
  \bibinfo{author}{\bibfnamefont{M.}~\bibnamefont{Deakyne}},
  \bibinfo{author}{\bibfnamefont{K.}~\bibnamefont{Lande}},
  \bibinfo{author}{\bibfnamefont{C.}~\bibnamefont{Lee}},
  \bibinfo{author}{\bibfnamefont{R.}~\bibnamefont{Steinberg}},
  \bibnamefont{et~al.}, \bibinfo{journal}{Phys.Rev.Lett.}
  \textbf{\bibinfo{volume}{47}}, \bibinfo{pages}{1507} (\bibinfo{year}{1981}).

\bibitem[{\citenamefont{Olive et~al.}(2014)}]{Agashe:2014kda}
\bibinfo{author}{\bibfnamefont{K.}~\bibnamefont{Olive}} \bibnamefont{et~al.}
  (\bibinfo{collaboration}{Particle Data Group}), \bibinfo{journal}{Chin.Phys.}
  \textbf{\bibinfo{volume}{C38}}, \bibinfo{pages}{090001}
  (\bibinfo{year}{2014}).

\bibitem[{\citenamefont{Silverman and Soni}(1981)}]{Silverman:1980ha}
\bibinfo{author}{\bibfnamefont{D.}~\bibnamefont{Silverman}} \bibnamefont{and}
  \bibinfo{author}{\bibfnamefont{A.}~\bibnamefont{Soni}},
  \bibinfo{journal}{Phys.Lett.} \textbf{\bibinfo{volume}{B100}},
  \bibinfo{pages}{131} (\bibinfo{year}{1981}).

\bibitem[{\citenamefont{Arnold et~al.}(2013)\citenamefont{Arnold, Fornal, and
  Wise}}]{Arnold:2013cva}
\bibinfo{author}{\bibfnamefont{J.~M.} \bibnamefont{Arnold}},
  \bibinfo{author}{\bibfnamefont{B.}~\bibnamefont{Fornal}}, \bibnamefont{and}
  \bibinfo{author}{\bibfnamefont{M.~B.} \bibnamefont{Wise}},
  \bibinfo{journal}{Phys.Rev.} \textbf{\bibinfo{volume}{D88}},
  \bibinfo{pages}{035009} (\bibinfo{year}{2013}), \eprint{1304.6119}.

\bibitem[{\citenamefont{Canetti et~al.}(2012)\citenamefont{Canetti, Drewes, and
  Shaposhnikov}}]{Canetti:2012zc}
\bibinfo{author}{\bibfnamefont{L.}~\bibnamefont{Canetti}},
  \bibinfo{author}{\bibfnamefont{M.}~\bibnamefont{Drewes}}, \bibnamefont{and}
  \bibinfo{author}{\bibfnamefont{M.}~\bibnamefont{Shaposhnikov}},
  \bibinfo{journal}{New J.Phys.} \textbf{\bibinfo{volume}{14}},
  \bibinfo{pages}{095012} (\bibinfo{year}{2012}), \eprint{1204.4186}.

\bibitem[{\citenamefont{Sakharov}(1967)}]{Sakharov:1967dj}
\bibinfo{author}{\bibfnamefont{A.}~\bibnamefont{Sakharov}},
  \bibinfo{journal}{Pisma Zh.Eksp.Teor.Fiz.} \textbf{\bibinfo{volume}{5}},
  \bibinfo{pages}{32} (\bibinfo{year}{1967}).

\bibitem[{\citenamefont{Berger et~al.}(1991)}]{Berger:1991fa}
\bibinfo{author}{\bibfnamefont{C.}~\bibnamefont{Berger}} \bibnamefont{et~al.}
  (\bibinfo{collaboration}{Frejus Collaboration}),
  \bibinfo{journal}{Phys.Lett.} \textbf{\bibinfo{volume}{B269}},
  \bibinfo{pages}{227} (\bibinfo{year}{1991}).

\bibitem[{\citenamefont{Araki et~al.}(2006)}]{Araki:2005jt}
\bibinfo{author}{\bibfnamefont{T.}~\bibnamefont{Araki}} \bibnamefont{et~al.}
  (\bibinfo{collaboration}{KamLAND Collaboration}),
  \bibinfo{journal}{Phys.Rev.Lett.} \textbf{\bibinfo{volume}{96}},
  \bibinfo{pages}{101802} (\bibinfo{year}{2006}), \eprint{hep-ex/0512059}.

\bibitem[{\citenamefont{Litos et~al.}(2014)\citenamefont{Litos, Abe, Hayato,
  Iida, Ikeda et~al.}}]{Litos:2014fxa}
\bibinfo{author}{\bibfnamefont{M.}~\bibnamefont{Litos}},
  \bibinfo{author}{\bibfnamefont{K.}~\bibnamefont{Abe}},
  \bibinfo{author}{\bibfnamefont{Y.}~\bibnamefont{Hayato}},
  \bibinfo{author}{\bibfnamefont{T.}~\bibnamefont{Iida}},
  \bibinfo{author}{\bibfnamefont{M.}~\bibnamefont{Ikeda}},
  \bibnamefont{et~al.}, \bibinfo{journal}{Phys.Rev.Lett.}
  \textbf{\bibinfo{volume}{112}}, \bibinfo{pages}{131803}
  (\bibinfo{year}{2014}).

\bibitem[{\citenamefont{Mohapatra and Senjanovic}(1982)}]{Mohapatra:1982aj}
\bibinfo{author}{\bibfnamefont{R.~N.} \bibnamefont{Mohapatra}}
  \bibnamefont{and}
  \bibinfo{author}{\bibfnamefont{G.}~\bibnamefont{Senjanovic}},
  \bibinfo{journal}{Phys.Rev.Lett.} \textbf{\bibinfo{volume}{49}},
  \bibinfo{pages}{7} (\bibinfo{year}{1982}).

\bibitem[{\citenamefont{Arnellos and Marciano}(1982)}]{Arnellos:1982nt}
\bibinfo{author}{\bibfnamefont{L.}~\bibnamefont{Arnellos}} \bibnamefont{and}
  \bibinfo{author}{\bibfnamefont{W.~J.} \bibnamefont{Marciano}},
  \bibinfo{journal}{Phys.Rev.Lett.} \textbf{\bibinfo{volume}{48}},
  \bibinfo{pages}{1708} (\bibinfo{year}{1982}).

\bibitem[{\citenamefont{Bryman}(2014)}]{Bryman:2014tta}
\bibinfo{author}{\bibfnamefont{D.}~\bibnamefont{Bryman}},
  \bibinfo{journal}{Phys.Lett.} \textbf{\bibinfo{volume}{B733}},
  \bibinfo{pages}{190} (\bibinfo{year}{2014}), \eprint{1404.7776}.

\bibitem[{\citenamefont{Fukuda et~al.}(2003)}]{Fukuda:2002uc}
\bibinfo{author}{\bibfnamefont{Y.}~\bibnamefont{Fukuda}} \bibnamefont{et~al.}
  (\bibinfo{collaboration}{Super-Kamiokande Collaboration}),
  \bibinfo{journal}{Nucl.Instrum.Meth.} \textbf{\bibinfo{volume}{A501}},
  \bibinfo{pages}{418} (\bibinfo{year}{2003}).

\bibitem[{\citenamefont{Abe et~al.}(2014{\natexlab{a}})}]{Abe:2013gga}
\bibinfo{author}{\bibfnamefont{K.}~\bibnamefont{Abe}} \bibnamefont{et~al.},
  \bibinfo{journal}{Nucl.Instrum.Meth.} \textbf{\bibinfo{volume}{A737}},
  \bibinfo{pages}{253} (\bibinfo{year}{2014}{\natexlab{a}}),
  \eprint{1307.0162}.

\bibitem[{\citenamefont{Abe et~al.}(2013)}]{Abe:2013lua}
\bibinfo{author}{\bibfnamefont{K.}~\bibnamefont{Abe}} \bibnamefont{et~al.}
  (\bibinfo{collaboration}{Super-Kamiokande Collaboration})
  (\bibinfo{year}{2013}), \eprint{1305.4391}.

\bibitem[{\citenamefont{Choi et~al.}(2015)}]{Choi:2015ara}
\bibinfo{author}{\bibfnamefont{K.}~\bibnamefont{Choi}} \bibnamefont{et~al.}
  (\bibinfo{collaboration}{Super-Kamiokande}),
  \bibinfo{journal}{Phys.Rev.Lett.} \textbf{\bibinfo{volume}{114}},
  \bibinfo{pages}{141301} (\bibinfo{year}{2015}), \eprint{1503.04858}.

\bibitem[{\citenamefont{Yamazaki and Akaishi}(2000)}]{Yamazaki:1999gz}
\bibinfo{author}{\bibfnamefont{T.}~\bibnamefont{Yamazaki}} \bibnamefont{and}
  \bibinfo{author}{\bibfnamefont{Y.}~\bibnamefont{Akaishi}},
  \bibinfo{journal}{Phys.Lett.} \textbf{\bibinfo{volume}{B453}},
  \bibinfo{pages}{1} (\bibinfo{year}{2000}).

\bibitem[{\citenamefont{Nishino et~al.}(2012)}]{Nishino:2012ipa}
\bibinfo{author}{\bibfnamefont{H.}~\bibnamefont{Nishino}} \bibnamefont{et~al.}
  (\bibinfo{collaboration}{Super-Kamiokande}), \bibinfo{journal}{Phys.Rev.}
  \textbf{\bibinfo{volume}{D85}}, \bibinfo{pages}{112001}
  (\bibinfo{year}{2012}), \eprint{1203.4030}.

\bibitem[{\citenamefont{Regis et~al.}(2012)}]{Regis:2012sn}
\bibinfo{author}{\bibfnamefont{C.}~\bibnamefont{Regis}} \bibnamefont{et~al.}
  (\bibinfo{collaboration}{Super-Kamiokande Collaboration}),
  \bibinfo{journal}{Phys.Rev.} \textbf{\bibinfo{volume}{D86}},
  \bibinfo{pages}{012006} (\bibinfo{year}{2012}), \eprint{1205.6538}.

\bibitem[{\citenamefont{Abe et~al.}(2014{\natexlab{b}})}]{Abe:2014mwa}
\bibinfo{author}{\bibfnamefont{K.}~\bibnamefont{Abe}} \bibnamefont{et~al.}
  (\bibinfo{collaboration}{Super-Kamiokande Collaboration}),
  \bibinfo{journal}{Phys.Rev.} \textbf{\bibinfo{volume}{D90}},
  \bibinfo{pages}{072005} (\bibinfo{year}{2014}{\natexlab{b}}),
  \eprint{1408.1195}.

\bibitem[{\citenamefont{Gustafson et~al.}(2015)}]{Gustafson:2015qyo}
\bibinfo{author}{\bibfnamefont{J.}~\bibnamefont{Gustafson}}
  \bibnamefont{et~al.} (\bibinfo{collaboration}{Super-Kamiokande}),
  \bibinfo{journal}{Phys.Rev.} \textbf{\bibinfo{volume}{D91}},
  \bibinfo{pages}{072009} (\bibinfo{year}{2015}), \eprint{1504.01041}.

\bibitem[{\citenamefont{Nakamura et~al.}(1976)\citenamefont{Nakamura,
  Hiramatsu, Kamae, Muramatsu, Izutsu et~al.}}]{Nakamura:1976mb}
\bibinfo{author}{\bibfnamefont{K.}~\bibnamefont{Nakamura}},
  \bibinfo{author}{\bibfnamefont{S.}~\bibnamefont{Hiramatsu}},
  \bibinfo{author}{\bibfnamefont{T.}~\bibnamefont{Kamae}},
  \bibinfo{author}{\bibfnamefont{H.}~\bibnamefont{Muramatsu}},
  \bibinfo{author}{\bibfnamefont{N.}~\bibnamefont{Izutsu}},
  \bibnamefont{et~al.}, \bibinfo{journal}{Nucl.Phys.}
  \textbf{\bibinfo{volume}{A268}}, \bibinfo{pages}{381} (\bibinfo{year}{1976}).

\bibitem[{\citenamefont{Brun et~al.}(1994)\citenamefont{Brun, Carminati, and
  Giani}}]{Brun:1994aa}
\bibinfo{author}{\bibfnamefont{R.}~\bibnamefont{Brun}},
  \bibinfo{author}{\bibfnamefont{F.}~\bibnamefont{Carminati}},
  \bibnamefont{and} \bibinfo{author}{\bibfnamefont{S.}~\bibnamefont{Giani}}
  (\bibinfo{year}{1994}).

\bibitem[{\citenamefont{Jadach et~al.}(1993)\citenamefont{Jadach, Was, Decker,
  and Kuhn}}]{Jadach:1993hs}
\bibinfo{author}{\bibfnamefont{S.}~\bibnamefont{Jadach}},
  \bibinfo{author}{\bibfnamefont{Z.}~\bibnamefont{Was}},
  \bibinfo{author}{\bibfnamefont{R.}~\bibnamefont{Decker}}, \bibnamefont{and}
  \bibinfo{author}{\bibfnamefont{J.~H.} \bibnamefont{Kuhn}},
  \bibinfo{journal}{Comput.Phys.Commun.} \textbf{\bibinfo{volume}{76}},
  \bibinfo{pages}{361} (\bibinfo{year}{1993}).

\bibitem[{\citenamefont{Chen and Takhistov}(2014)}]{Chen:2014ifa}
\bibinfo{author}{\bibfnamefont{M.-C.} \bibnamefont{Chen}} \bibnamefont{and}
  \bibinfo{author}{\bibfnamefont{V.}~\bibnamefont{Takhistov}},
  \bibinfo{journal}{Phys.Rev.} \textbf{\bibinfo{volume}{D89}},
  \bibinfo{pages}{095003} (\bibinfo{year}{2014}), \eprint{1402.7360}.

\bibitem[{\citenamefont{Honda et~al.}(2007)\citenamefont{Honda, Kajita,
  Kasahara, Midorikawa, and Sanuki}}]{Honda:2006qj}
\bibinfo{author}{\bibfnamefont{M.}~\bibnamefont{Honda}},
  \bibinfo{author}{\bibfnamefont{T.}~\bibnamefont{Kajita}},
  \bibinfo{author}{\bibfnamefont{K.}~\bibnamefont{Kasahara}},
  \bibinfo{author}{\bibfnamefont{S.}~\bibnamefont{Midorikawa}},
  \bibnamefont{and} \bibinfo{author}{\bibfnamefont{T.}~\bibnamefont{Sanuki}},
  \bibinfo{journal}{Phys.Rev.} \textbf{\bibinfo{volume}{D75}},
  \bibinfo{pages}{043006} (\bibinfo{year}{2007}), \eprint{astro-ph/0611418}.

\bibitem[{\citenamefont{Hayato}(2002)}]{Hayato:2002sd}
\bibinfo{author}{\bibfnamefont{Y.}~\bibnamefont{Hayato}},
  \bibinfo{journal}{Nucl.Phys.Proc.Suppl.} \textbf{\bibinfo{volume}{112}},
  \bibinfo{pages}{171} (\bibinfo{year}{2002}).

\bibitem[{\citenamefont{Abe et~al.}(2015)}]{Abe:2014gda}
\bibinfo{author}{\bibfnamefont{K.}~\bibnamefont{Abe}} \bibnamefont{et~al.}
  (\bibinfo{collaboration}{Super-Kamiokande}), \bibinfo{journal}{Phys. Rev.}
  \textbf{\bibinfo{volume}{D91}}, \bibinfo{pages}{052019}
  (\bibinfo{year}{2015}), \eprint{1410.2008}.

\bibitem[{\citenamefont{Shiozawa}(1999)}]{Shiozawa:1999sd}
\bibinfo{author}{\bibfnamefont{M.}~\bibnamefont{Shiozawa}}
  (\bibinfo{collaboration}{Super-Kamiokande Collaboration}),
  \bibinfo{journal}{Nucl.Instrum.Meth.} \textbf{\bibinfo{volume}{A433}},
  \bibinfo{pages}{240} (\bibinfo{year}{1999}).

\bibitem[{\citenamefont{Wendell et~al.}(2010)}]{Wendell:2010md}
\bibinfo{author}{\bibfnamefont{R.}~\bibnamefont{Wendell}} \bibnamefont{et~al.}
  (\bibinfo{collaboration}{Super-Kamiokande Collaboration}),
  \bibinfo{journal}{Phys.Rev.} \textbf{\bibinfo{volume}{D81}},
  \bibinfo{pages}{092004} (\bibinfo{year}{2010}), \eprint{1002.3471}.

\end{thebibliography}

\end{document}